\documentclass[aps,pra,superscriptaddress,amsmath,amsfonts,amssymb,floatfix,nofootinbib]{revtex4}
\usepackage{enumerate}
\usepackage{mathtools}
\usepackage{amsmath,amsthm}
\usepackage{graphicx,algorithm,algorithmic}
\usepackage{epsfig}
\usepackage{sidecap}
\usepackage{hyperref,stackrel}
\usepackage[normalem]{ulem}
\usepackage{color}
\usepackage{enumerate}
\usepackage{bm}
\usepackage[utf8]{inputenc}
\usepackage{amsmath}
\usepackage{amsfonts}
\usepackage{amssymb,wasysym}
\usepackage{graphicx}
\usepackage{enumitem}
\usepackage{lmodern}
\usepackage [english]{babel}
\usepackage [autostyle, english = american]{csquotes}
\MakeOuterQuote{"}
\newtheorem{definition}{Definition}

\begin{document}
\title{Quantum anonymous veto: A set of new protocols}

\author{Sandeep Mishra} \thanks{sandeep.mtec@gmail.com }
\affiliation{Jaypee Institute of Information Technology, A-10, Sector-62, Noida, UP-201309, India}

\author{Kishore Thapliyal} \thanks{kishore.thapliyal@upol.cz}
\affiliation{Joint Laboratory of Optics of
Palack\'{y} University and Institute of Physics of CAS, Faculty of Science,
Palack\'{y} University, 17. listopadu 12, 771 46 Olomouc, Czech Republic}

\author{Abhishek Parakh} \thanks{aparakh@unomaha.edu}
\affiliation{University of Nebraska, Omaha, USA }

\author{Anirban Pathak} \thanks{anirban.pathak@gmail.com}
\affiliation{Jaypee Institute of Information Technology, A-10, Sector-62, Noida, UP-201309, India}

\begin{abstract}
We propose a set of protocols for quantum anonymous veto (QAV) broadly categorized under the probabilistic, iterative, and deterministic schemes. The schemes are based upon different types of quantum resources. Specifically, they may be viewed as single photon-based, bipartite and multipartite entangled states-based, orthogonal state-based and conjugate coding-based. The set of the proposed schemes is analyzed for all the requirements of a valid QAV scheme (e.g., privacy, verifiability, robustness, binding, eligibility and correctness). 
The proposed schemes are observed to be more efficient in comparison to the existing QAV schemes and robust up to the moderate decoherence rate. In addition, a trade-off between correctness and robustness of the probabilistic QAV schemes is observed. 
Further, the multipartite dense coding based determinsitic QAV scheme is most efficient scheme among the set of schemes proposed here. A bipartite entanglement based iterative scheme employing dense coding is yet another efficient and practical scheme. The intrinsic connections between dining cryptographer-net with anonymous veto-net is also explored in the process of designing new protocols.

\end{abstract}

\maketitle

\section{Introduction}

Everyday humans have to deal with conflicting issues which demand making some choices and in modern societies, voting is an integral part of those decision making processes. In simple words, everyone having a stake exercises the possible choice of option and finally a decision is arrived at. In most of the cases, the outcome of the voting is based on the majority voting outcome. But sometimes there may be situations where a split outcome is not desireable as the consequences may be too high. So, it is required that any decision that is taken is arrived at by a consensus only. For instance, a verdict for the capital punishment cannot be adjudged only on the majority view of the judges as no judicial system is perfect, and the life of the accused cannot be revived in view of evidences to acquit him posthumously. In such a case, capital punishment is overturned even if one of the judge dissents. Similarly, in big corporations, shareholders may like to exercise their votes before making some crucial decisions. Some of the stakeholders may collude to influence the decision to sabotage the stakes of their rivals.  Such a situation naturally desires a process in which the decision is made by consensus. The most glaring example is the United Nations security council resolutions, in which a proposal is rejected at once if one or more of the P5 countries exercise(s) its \emph{veto} power. Therefore, veto empowers a voter in the voting process to reject a proposal unilaterally.
In other words, a proposal is rejected even if one of the voters does not approve the proposal, and thus a decision can only be made unanimously. Usually in the veto scheme, the group of voters is limited and no one would like to reveal their identity after exercising the veto as it may have some repercussions. Therefore, the useful information is only a single bit, i.e., whether the decision is made by consensus or not decision has been reached (which means someone vetoed the proposal in the latter case).

With the advent of  quantum enabled technologies, certain tasks are achievable, which were not possible otherwise with the use of classical resources only; e.g., the current classical and post-quantum cryptographic systems exploit the mathematical complexity associated with the process of solving certain problems assumed to be hard on classical computers \cite{shenoy2017quantum}. However, many such cryptographic systems are vulnerable to a scalable quantum computer, which can implement quantum algorithms to solve these respective problems efficiently \cite{gisin2002quantum,shenoy2017quantum}. In contrast, with the advent of quantum cryptographic schemes, such as BB84 \cite{bennett1984quantum} and E91 \cite{ekert1991} quantum key distribution (QKD) protocols, quantum mechanics equipped us with the feasibility of unconditional secure communication. Here, unconditional security corresponds to the fact that it's based on the laws of physics governed by quantum mechanics and is not conditioned on the computational power available to an adversary. 
This motivated a host of new protocols for secure quantum communication and/or quantum computation. Specifically, secure quantum computation comprises the features of both computation and communication as it enables us to compute a multi-variable function, with each input provided by different individuals, in such a way that the inputs are not disclosed. Quantum solutions for the tasks, such as secure multi party computation \cite{qsmpc1,qsmpc2}, private comparison \cite{qpc1,qpc2}, auctions \cite{qa1,qa2,sharma2017}, provide examples of situations where quantum advantage is obtained in the field of secure computation. This also inspired the use of quantum resources in the field of voting as it requires features such as anonymity, verifiability and security from tampering. In 2006, the first set of quantum voting protocols was proposed using quantum entangled states \cite{hillery2006, vaccaro2007}. Since then a large number of new protocols for anonymous voting have been designed, but an unconditionally secure and practical quantum voting protocol has remained elusive till this date \cite{thapliyal2017qv,wang2016qv,jiang2012qv,hillery2011qv,xue2017qv}. {More recently, there has been a heightened interest in the quantum anonymous voting protocols with a flurry of papers  \cite{sun2019qv,jiang2020quantum,liu2021quantum,
wang2020quantum,wang2021quantum,du2021secure,shi2021anonymous,sekga2021quantum,
liu2021novel,li2020novel,li2021quantum,zhang2020secure,joy2020implementation,wang2021}. }
These voting schemes can be classified in different categories on the basis of required quantum resources, nature of ballots, number of candidates, conditions to be satisfied, and so on. 

One such interesting voting scheme is an anonymous veto (AV) protocol, which has not been studied much. Specifically, Rahman and Kar introduced the idea of quantum solution for AV using GHZ states to implement privacy while casting a veto \cite{rahaman1}. The idea was to explore the interconnections between the dinning cryptographers (DC) net problem and AV net problem \citep{ding2020}. The idea of DC nets was introduced in 1988 to illustrate a scheme in which parties can send messages with cryptographically secure non-traceability \cite{chaum1988}. Specifically, DC nets are based on establishing the secret keys (between every pair of the parties) as one of the primitives. Since then DC nets have been used as one of the possible ways to implement anonymous broadcasting of the messages. In 2021, analogous to \cite{rahaman1}, a new quantum AV (QAV) protocol based on $n$-party GHZ states was proposed with a proof of principle experiment on quantum computer placed on cloud by the IBM Corporation for four voters \cite{wang2021}.   However, the scheme in \cite{wang2021} is neither practical nor efficient, which motivated us to propose some QAV protocols based on different quantum states. Specifically, we propose here the protocols for QAV scheme using single photon, Bell state, GHZ and cluster state, which can be implemented between voters equipped with different kinds of quantum resources. In the present work, we have also been able to show the intrinsic connections between the AV nets and DC nets. In view of some recent works, we expect the applications of our QAV protocols for the implementation of sealed bid auctions \cite{bag2019seal}.

The rest of the paper is structured as follows. In section \ref{sec:basic}, we introduce the basic ideas and nomenclature used in the present work. Subsequently, we begin with reviewing the existing schemes of QAV with their limitations in section \ref{sec:exSc} followed by our new set of QAV schemes in section \ref{sec:newSc}. We present the security and efficiency analysis of the proposed schemes in sections \ref{sec:SE}. And finally, we summarize the results in section \ref{sec:conc}.

\section{Basic notations and definitions}\label{sec:basic}

\begin{definition}
An AV protocol of $n$ voters returns $\mathcal{V}_n=0$ if all the voters support the proposal and $\mathcal{V}_n=1$ otherwise. In other words, an $n$ input function $\mathcal{V}_n\in\{0,1\}$ is computed as
\begin{equation}
\mathcal{V}_n = \lor_{i} \mathcal{W}_i=\Bigg\{\begin{array}{l}
0 \quad \mathrm{iff} \, \mathcal{W}_i=0\, \forall i \\
1 \quad \mathrm{otherwise}
\end{array}, 
\end{equation}
where the $i^{th}$ input $\mathcal{W}_i\in\{0,1\}$ is supplied by the $i^{th}$ voter, and the logical OR operation $\lor_{i}$ performed over all the $i$ inputs returns 0 only when all the inputs are 0. Thus,
{$\mathcal{V}_n=0$ or 1 provides whether $k=0$ or $k\neq 0$ number of voters veto the proposal among all the $n$ voters, respectively.}
\end{definition}

\subsection{Requirements of anonymous veto protocol}

Any AV protocol must conform to the following requirements in order to be classified as a good voting scheme \cite{hillery2006, vaccaro2007, schneier1996}.  

\begin{description}
\item[Eligibility:] No one except the authorized voters shall be allowed to vote.

\item[Privacy:] It means that nobody except the voter should be able to know how a particular voter has voted. 

\item[Binding:] No one (including the voter himself) can change the vote $\mathcal{W}_i$ after its submission. 

\item[Correctness:] If the adversary is passive, then the result bit $\mathcal{V}_n=0 {\iff} \mathcal{W}_i=0\, \forall i $ is generated. In other words, it means that after faithfully following the protocol, one is able to successfully detect a veto or unanimous agreement with probability $1$. 

\item[Verifiability:] All the participants can verify the result $\mathcal{V}_n$.

\item[Robustness:] If the adversary is passive, then the result bit $\mathcal{V}_n=i\,\forall i\in\{0,1\}$  is generated.  It means that the system obtains the result if adversary is passive, i.e. under the effect of the noise in the systems. 

\end{description}

\subsection{Authentication using quantum digital signatures}
The first and foremost thing in any voting scheme is to provide a mechanism to establish that only genuine and eligible voters are allowed to take part in the voting process.  It can also be referred to as pre-voting stage. Various classical authentication schemes are available for the verification of the authenticity of a voter, but the security of such schemes is usually based on computational complexity only. Here, we will be using a quantum digital signature scheme based on BB84 states as proposed in \cite{wallden2015quantum} to verify the authenticity of the eligible voters. Suppose, there is a trusted central authority (CA) who will verify the credentials of the voters. After verification, the voter is registered and asked to generate his digital signature. The voter then sends a sufficiently long sequence of BB84 states ($\vert 0 \rangle, \vert 1 \rangle, \vert + \rangle, \vert - \rangle$) to CA. CA receives the states and measures them randomly in the computational basis ($\vert 0 \rangle, \vert 1 \rangle$)  or Hadamard basis ($\vert + \rangle, \vert - \rangle$). After the measurement, CA eliminates one of the BB84 states that the voter must have never sent. For instance, if CA's measurement outcome is $\vert 0 \rangle$ then it infers the voter has not sent $\vert 1 \rangle$. The measurement outcomes of CA thus form an eliminated signature of the voter. During the voting stage for the purpose of  authentication, each voter will reveal to CA the choice of BB84 states (digital signature) that they have sent in the pre-voting stage. CA will then verify the digital signature with the eliminated signature of the voter and if the number of mismatches is lesser than a particular threshold then the authentication of the voter is validated. After authentication, the voter is allowed to take part in the voting process for casting the vote.

\subsection{Decoy state based eavesdropping checking techniques}

In quantum cryptography, security is achieved by obtaining an upper bound on the information accessible to Eve by checking the error rates in the transmission of qubits. This is based on the fact that any eavesdropping attempt leaves detectable traces at the receiver's end. Therefore, some verification qubits, known as decoy qubits \cite{sharma2016verification}, are inserted randomly in the string of qubits before transmission to be used for parameter estimation. In other words, for the secure transmission of $t$ qubits through a channel accessible to Eve an additional $\delta t$ decoy qubits are inserted. The factor of $\delta > 0$ is decided to achieve the desired level of security. For example, it is shown that the probability of obtaining more than $(\Delta+\epsilon) t\,$ errors in the transmitted qubits (such that $\Delta>0,\epsilon>0$)for $\Delta \delta t$ errors on the decoy qubits is asymptotically less than $\exp[-O(\epsilon^2 t)]$ for $\delta=1$ \cite{nielsen}.

Decoy state based eavesdropping checking techniques may be broadly categorized on the basis of nature of verification qubits as follows (\cite{sharma2016verification} and references therein).
\begin{enumerate}
\item \emph{BB84 subroutine}: The set of BB84 states is inserted by the sender in the travel qubits randomly and the receiver measures them after the sender informs him the position and the basis chosen to prepare the state. All the errors in the measurement outcome (including due to transmission noise) when compared with the state prepared are attributed to the eavesdropping attempt. The name suggests that security comes from Eve's inability to measure a quantum state in mutually unbiased bases without leaving detectable traces.
\item \emph{GV subroutine}: Multiple copies of one of the entangled states (say a Bell state) are used as decoy states while the position of the entangled particles is kept secret while transmission. This geographical separation of entangled qubits restricts Eve from measuring the state in the publicly known basis. An eavesdropping attempts leads to entanglement swapping and detectable traces as errors in the receiver's port.  
\end{enumerate}
It would be worth mentioning here that in semiquantum cryptography a two-way communication of the decoy qubits is involved as a classical user reflects all the decoy qubits as he is restricted to measure in the computational basis only. Thus, in what follows, a secure transmission of qubits is performed using decoy state based eavesdropping checking technique.

\section{Existing protocols and their limitations}\label{sec:exSc}

We briefly review two quantum anonymous veto protocols, and mention some of their limitations. Specifically, we summarize a quantum anonymous veto protocol proposed by Rahman and Kar (RK) referred to as RKQAV protocol \cite{rahaman1}, which motivated Wang et al.'s scheme \cite{wang2021} referred to as WQAV protocol.

\subsection{Iterative QAV protocol: RKQAV protocol}
Rahman and Kar \cite{rahaman1} proposed RKQAV protocol using the properties of multiple copies of $n$-qubit GHZ states \cite{ghz1989} shared among $n$ voters $V_i\, \forall \, i\in \{0,1,\dots,n-1 \}$. Without loss of generality, we assume that $0\leq k\leq n$ voters veto the proposal. The steps involved in the protocol based on generalization of dining cryptographers protocol can be briefly mentioned as follows: 

\begin{description}
\item [{RKQAV~1:}] $l\, (l \ge 2)$ ordered copies of $n$-qubit GHZ states  
 \begin{equation}\label{eq:GHZ}
 \vert \chi\rangle_j = \frac{1}{\sqrt{2}} (\vert 0 \rangle^{\otimes n} + \vert 1 \rangle^{\otimes n})\,\forall\, j= 0,1,\dots,l-1 
 \end{equation}
are shared among the $n$ voters in such a manner that each voter receives one qubit of each of the GHZ states\footnote{In Ref.~\cite{rahaman1}, it is not explicitly mentioned who prepares and shares them among the voters. For the sake of completeness, we may assume here one of the voters or a trusted third party prepares and shares it among them.}. They check the shared correlations by verifying GHZ-type paradox.
 
\item [{RKQAV~2:}] All the voters select one of the shared $l$ copies of the GHZ state randomly for encoding (say $m^{th}$ copy). $k$ voters perform a unitary operation $\sigma_z$ on his qubit of the $m^{th}$ GHZ state while the rest of the $n-k$ voters do nothing. 

\item [{RKQAV~3:}] If the final state of the $m^{th}$ GHZ state remains unchanged, i.e., they obtain $\vert \chi \rangle_m$ on GHZ measurement, it corresponds to either $k=0$ or $k$ is non-zero even number. However, in case of odd $k$, the final joint state will be orthogonal to the initial state, i.e.,
\begin{equation}
 \vert \chi\rangle^{\perp}_m = \frac{1}{\sqrt{2}} (\vert 0 \rangle^{\otimes n} - \vert 1 \rangle^{\otimes n}).
 \end{equation}
This allows the voters to distinguish whether an odd number of voters have vetoed the proposal (for $\vert \chi\rangle^{\perp}_m$) or an inconclusive outcome (for $\vert \chi\rangle_m$) is obtained. For the final joint state $\vert \chi\rangle_m$ no conclusion can be made as the state can be obtained in following cases:  (a) all are in `favour' of the proposal or (b) an even number of voters have vetoed the proposal.\\
In case of conclusive outcome $\vert \chi\rangle^{\perp}_m$ they have accomplished the desired task, while for an inconclusive outcome $\vert \chi\rangle_m$ they proceed with the protocol. To distinguish between the two cases of an inconclusive outcome (a) $k=0$ and (b) $k$ non-zero even number, they repeat the {next step} for a few iterations. 
 
\item [{RKQAV~4:}] In the $t^{th}$ (for $t\geq 1$) iteration\footnote{Interestingly, RKQAV 3 can be viewed as $(t=0)$th iteration.}, $k$ voters apply a unitary operation $ \sigma_{z}(t) = \begin{pmatrix}
1 & 0 \\
0 & \exp\left({i \pi}{2^{-t}}\right)
\end{pmatrix}$ to convey they are against the proposal, while $n-k$ do nothing, on their respective qubits of the randomly chosen GHZ state among the remaining $l-t$ copies.\\
In each iteration, if their measurement outcome results $\vert \chi\rangle$ then $2^{-t}k$ is even (including zero), while the final state $\vert \chi\rangle^{\perp}$ corresponds an odd value of $2^{-t}k$.\\
They truncate this iteration until either the measurement outcome is $\vert \chi\rangle^{\perp}$ or they get $\vert \chi\rangle$ for $k=0$ conclusively. 

\item [{RKQAV~5:}] Since, the total number of voters is finite and after every round we eliminate half of the possibilities, so after a finite number of steps one can detect whether there is any unanimity `in favor' of the decision or at least one voter has vetoed the proposal.

\end{description}

RKQAV protocol provided an initial idea to implement AV using quantum states, but this protocol was not mature (due to a large number of loopholes) to be implemented with real systems. Specifically, in the original proposal of RK, it was not mentioned who is responsible for the generation and distribution of the GHZ states. Nothing was stated about how and who will have the responsibility to distinguish between the GHZ states $\vert \chi\rangle$ and $\vert \chi\rangle^{\perp}$. Further, no elaborate security analysis of the protocol with respect to an ideal quantum voting protocol was reported. {Additionally, the implementation of the protocol requires a maximum of $\sim(1+\log_2 n)$ number of iterations to yield a conclusive outcome $\mathcal{V}_n$, so we refer to this scheme as iterative QAV protocol.}

\subsection{Probabilistic QAV protocol: WQAV protocol}

Wang et al. \cite{wang2021} further improved the RKQAV protocol and presented a new and mature mechanism (WQAV). Similar to RKQAV, this protocol utilizes the GHZ state, while allows measurement by voters in the computational ($\{\vert 0 \rangle, \vert 1 \rangle\}$) and diagonal ($\{\vert + \rangle, \vert - \rangle\}$) basis in addition of single qubit unitary operations, i.e. $\sigma_{z}$ and Hadamard gate. In this case, the quantum voting network consists of $n$ voters $V_i\, \forall \, i\in \{0,1,\dots,n-1 \}$ which is controlled by a {semi-honest} central authority (CA).

The steps involved in the protocol can be briefly described as follows:

\begin{description}

\item [{WQAV~1:}] CA authenticates every voter and shares a binary key $B_i = \{b_{ij}\}_{j=0,1, \dots,l-1}$ of $l$-bits using any QKD protocol, such as BB84 protocol \cite{bennett1984quantum}, with the $i^{th}$ voter $V_i$.

\item  [{WQAV~2:}] CA distributes $l$ ordered copies of $n$-qubit GHZ states (\ref{eq:GHZ}) in a secure manner such that each voter receives a qubit of entangled states.
 
\item [{WQAV~3:}] Only if each voter finds the error rate in the eavesdropping check below the threshold error, they proceed with the protocol and each voter $V_i$ possesses $l$ ordered particles given by $S_i=\{s^i_{j}\}_{j=0,1,\dots,l-1}$.

\item [{WQAV~4:}] Each voter encodes their voting information. Specifically, $k$ voters perform a local phase flip gate $\sigma_{z}$ to every particle $s^{i}_{j}$ with probability $1/2$, while the rest of the $n-k$ voters do nothing. 

\item [{WQAV~5:}] Voter $V_i$ applies a Hadamard operation  on his set of particles $S_i$ before measuring them in the computational basis. This will result in the generation of $l$ ordered intermediate data sequence $T_i = \{t_{ij}\}_{j=0,1,\dots,l-1}$ for each voter $V_i$.

\item [{WQAV~6:}] Every voter $V_i$ then transmits his data sequence $T_i$ to CA via the use of the shared secret  key $B_i$ with CA. Specifically, the data sequence received by CA from each voter is $Y_{i}= T_i\bigoplus B_i = \{y_{ij} = (t_{ij} +b_{ij}) \,\mathrm{mod}\,  2\}_{j=0,1,\dots,l-1} $.

\item [{WQAV~7:}]  CA then calculates $\{R_j\}_{j=0,1,\dots,l-1}= Y_i\bigoplus B_i$ with 
\begin{equation}
R_j= \sum^{n-1}_{i=0} (y_{ij} + b_{ij} )\, \mathrm{mod}\, 2.
\end{equation}

\item [{WQAV~8:}] If CA gets at least one $R_j \ne 0$, then at least one voter has vetoed the proposal. In other words, $\mathcal{V}_n=0 \Leftrightarrow R_j=0\,\forall j$. The probability for CA to successfully detect a veto is given by $\left(1-{2^{-l}}\right)$. This is due to the fact that if $k=0$ then the number of outcomes "1'' in the sequence $\{t_{0,j},t_{1,j}, \dots, t_{n-1,j} \}$  for every $j=0,1,\dots,l-1$ is even. However, if at least one voter has vetoed the protocol, then the probability of successfully getting  "1'' for each $R_j$ is 1/2.

\end{description}

WQAV protocol has elaborately discussed  the  mechanism for secure distribution of shared GHZ states and have discussed the security analysis. Further, it has been proved that this protocol satisfies the essential requirements of any secure voting protocol. The disadvantage of WQAV is that it requires a large amount of quantum resources. For example, the CA has to generate $l$ bit keys with all the voters using QKD/quantum key agreement (QKA) protocol which will require additional resources. Further, the $n$ party GHZ state is difficult to generate and maintain, and here $l$ copies of such a state are required. Only ideal cases have been considered while a practical protocol should be robust against noise. Specifically, due to the effect of noise the correlations in GHZ states will reduce, which may lead to false veto or vice versa. The scheme, although achieving the task in a single iteration, remains a probabilistic QAV protocol unless $l$ is large. Further, there is no discussion on the  possibilities of improving the efficiency or robustness of the protocol. 

\section{New protocols for quantum veto}\label{sec:newSc}

The field for development of unconditionally secure AV protocol using quantum resources is still at a nascent stage. In the following, we propose a few protocols in order to implement AV scheme using the optimal utilization of quantum resources and perform their security analysis. Specifically, we divide the proposed schemes in three broad categories: (i) probabilistic QAV,  (ii) iterative QAV, and (iii) deterministic QAV protocols. 

\begin{figure}[tb]
\includegraphics[width=\textwidth]{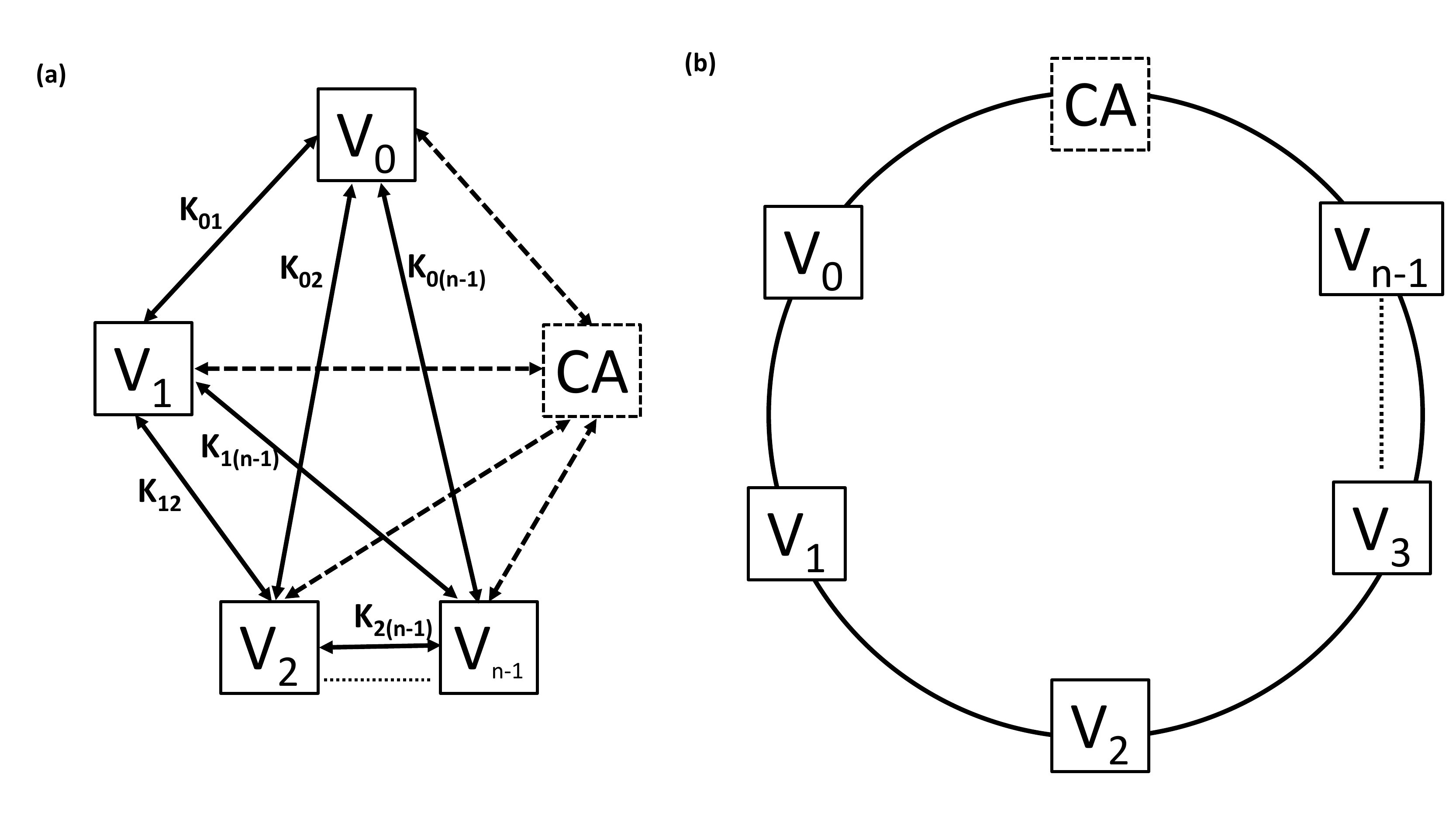}
\caption{Schematic arrangement of voters in a quantum voting network with $n$ voters in a (a) complete graph and (b) circular structures. A detailed description is given in the text. } 
\label{fig1}
\end{figure}

\subsection{Probabilistic QAV protocols}

We assume that we have a quantum voting network with a {semi-honest} central authority CA (unless stated otherwise) and $n$ voters $V_i$s. Without loss of generality, we may assume that $0\leq k\leq n$ voters veto the proposal.

\subsubsection{QAV-1: Quantum key distribution/key agreement based QAV protocol}

A probabilistic QAV protocol, which uses the complete graph structure (Fig. \ref{fig1} (a)) can be proposed using any of the existing protocols for QKD or QKA. Specifically, the dashed lines in Fig. \ref{fig1} (a) in the tree structure represent classical communication, while the smooth lines correspond to the quantum transmission. The steps involved in this protocol (QAV-1) are as follows:

\begin{description}

\item [{Step~1.1}] CA verifies the validity of every voter $V_i$ {by using the method of quantum digital signature described in Section \ref{sec:basic}}. 

\item [{Step~1.2}] Voter $V_i$ generates a $l$ bit symmetric key $V^{ij}=\{ v^{ij}_{l'} \}_{j\neq i}\, \forall \,i,j=0,1,\dots,n-1$ with $v^{ij}_{l'} \in \{ 0,1\}$, and $l'=0,1,\dots,l-1$ with voter $V_{j\neq i}$ using a QKD or QKA protocol. 

\item [{Step~1.3}] Voter $V_i$ computes his sequence $V^{i}_{l'}=\sum^{n-1}_{j=0,j\neq i} v^{ij}_{l'}\, \mathrm{mod}\,2$ for every bit value of $l'$. All $n-k$ voters in favour of the proposal announce $V^{i}_{l'}$, while $k$ voters either broadcast $V^{i}_{l'}$ or apply a not gate to  $V^{i}_{l'}$ before broadcasting with an equal probability.

\item [{Step~1.4}] CA will compute the sequence $S_{l'}= \sum^{n-1}_{i=0} V^{i}_{l'} \mathrm{mod}2$ for every $l'=1,2,\dots,l$.

\item [{Step~1.5}] Thus, $\mathcal{V}_n=0 \Leftrightarrow S_{l'}=0\,\forall l'$. If at least one of the voter has used his veto power, then CA will get at least one of the $S_{l'} \ne 0$ and the success probability of detecting a veto is given by $1-2^{-l}$.  

\end{description}

{An important point to be mentioned here is that the role of CA in this protocol is only for the authentication of the eligible voters. Thus, this protocol can be implemented without CA if  the voters have the ability to authenticate each other. The protocol described above has a similar structure to  that of the DC net as this also requires to establish symmetric keys between all the pairs of voters before the voters can start casting their votes.} Further, in principle, all the voters can use different types of quantum resources to share the keys among them, such as using single photon based \cite{bennett1984quantum,bennett1992quantum}, orthogonal state based \cite{goldenberg1995quantum}, entangled state based \cite{ekert1991,bennett1992quantumcryptography}, counterfactual \cite{noh2009counterfactual}, semi-quantum \cite{PhysRevLett.99.140501}, continuous variable \cite{srikara2020continuous} QKD and/or QKA \cite{shukla2014protocols,shukla2017semi}, which will give the corresponding flavor to the proposed QAV scheme. Thus, this protocol is general in nature and actually represent a family of protocols which can be reduced to a specific protocol by choice of specific scheme(s) of QKD/QKA. The above fact is established by providing three protocols (cf. QAV-2-QAV-4), which can be viewed as specific protocols reduced from a more general protocol described here. In principle, each pair of voters can choose different QKD/QKA scheme independently, which will result in a hybrid QAV scheme.

\subsubsection{QAV-2: Bell state based probabilistic QAV protocol}

This protocol too involves the arrangement of voters in a  tree structure as shown in Fig. \ref{fig1} (a) and shows that the same task performed by WQAV protocol can be accomplished using solely bipartite entanglement, which is much easier to be produced and maintained. Specifically, the dashed lines in Fig. \ref{fig1} (a) in the tree structure represent quantum communication, while the smooth lines correspond to the shared entanglement.
The steps involved in this protocol (QAV-2) are as follows:

\begin{description}

\item [{Step~2.1}] Same as Step 1.1 of QAV-1. 

\item [{Step~2.2}] CA securely distributes $l$ Bell states $\vert \phi \rangle = \frac{1}{\sqrt{2}} (\vert 00 \rangle + \vert 11 \rangle) $ for each pair of the voters in a secure manner, i.e., voter $V_i$ shares $l$ copies of $\vert \phi \rangle$ states with each of the $n-1$ other voters. 

\item [{Step~2.3}] Same as WQAV 3, but here $V_i$ possesses $l$ strings of $n-1$ ordered particles given by $\{s^{ij}_{l'} \}_{j\neq i}\, \forall \,i,j=0,1,\dots,n-1$ with $l'=0,1,\dots,l-1$.

\item[{Step~2.4}] Same as WQAV 4.

\item [{Step~2.5}] Similar to WQAV 5, Voter $V_i$ measures all the qubits {in the computational basis} after performing a Hadamard operation and obtains $l$ binary sequences of $n-1$ bits given by $\{ v^{ij}_{l'} \}_{j\neq i} \, \forall \,i,j=0,1,\dots,n-1$ with $v^{ij}_{l'} \in \{ 0,1\}$. 

\item [{Step~2.6}] Similar to WQAV 6, Voter $V_i$ computes his sequence $V^{i}_{l'}=\sum^{n-1}_{j=0,j\neq i} v^{ij}_{l'}\, \mathrm{mod}\,2$ for every bit value of $l'$. Subsequently, Voter $V_i$ broadcasts his data sequence $V^{i}_{l'}$.

\item [{Step~2.7}] Similar to WQAV 7, CA (in principle, all the voters) calculates the $\{R_{l'}\}$ with $R_{l'}= \sum^{n-1}_{i=0} V^{i}_{l'}\, \mathrm{mod}\, 2$.\\
Thus, $\mathcal{V}_n=0 \Leftrightarrow R_{l'}=0\,\forall l'$ and if CA gets at least one $R_{l'} \neq 0$, then $k>0$.
Similar to WQAV and QAV-1, the probability for CA to successfully detect a veto is given by $\left(1-2^{-l}\right)$. 

\end{description}

 The protocol described above has a very close resemblance with the DC-net problem with regards
to its advantages and disadvantages. The main disadvantage of this protocol is that we need to a minimum of $l \times ^nC_{2}$ copies of Bell states which would consume a considerable amount of quantum
 resources. However, it does address the practical challenges in preparation of multipartite entangled state and robustness of WQAV protocol.

\subsubsection{QAV-3 and QAV-4: Alternative protocols of QAV for voters with limited resources}

QAV-1 can be easily modified to be implemented by the parties with unique quantum resources/abilities. For example, using an orthogonal state based QKA scheme \cite{shukla2014protocols} between each pair of voters completely orthogonal state based QAV protocol is proposed here. The feasibility of GV protocol \cite{goldenberg1995quantum} established that the unconditional security of quantum cryptography can be achieved using solely orthogonal states. This is achieved by making the basis of preparation of the state inaccessible to the eavesdropper by using temporal or geographical splitting of different quantum pieces.  
We briefly describe the modification in Step 1.2 of QAV-1 to design an orthogonal state based protocol (QAV-3), while the rest of the steps remain unchanged.

\begin{description}

\item [{Step~3.2.1}] $V_i$ generates $\frac{l}{2}$ Bell states $\vert \phi \rangle$ to be shared with voter $V_{j\neq i}$ and forms two ordered sequences of all the first qubits $H^{ij}=\{h^{ij}_{l'} \}_{j\neq i}$ and second qubits $T^{ij}=\{t^{ij}_{l'} \}_{j\neq i}$ with $l'=0,1,\dots,\frac{l}{2}-1$. Similarly, all the set of voters $i,j=0,1,\dots,n-1$ prepare the home and travel sequences $H^{ij}$ and $T^{ij}$, respectively.\\
All the voters $V_i$ prepare a random $l$-bit sequence $K^{ij}= \{k^{ij}_{l'} \}_{j\neq i}$, where $k^{ij}_{l'} \in \{0,1 \} $, to share the rest of the voters $V_{j\neq i}$.

\item [{Step~3.2.2}] $V_i$ concatenates some Bell states to $T^{ij}$ and applies a permutation operator to the enlarged travel sequence $T^{ij}_1$ before sending them to $V_{j}$. $V_i$ reveals the correct order of sequence only after he receives an authenticated acknowledgment of the receipt of qubits from $V_{j}$.

\item [{Step~3.2.3}] If the error rate in the eavesdropping checking is below the pre-determined threshold value, $V_{j}$ applies the unitary operation $I$, $\sigma_x$, $i\sigma_y$, and $\sigma_z$ to the sequence $T^{ij}$ in order to encode 00, 01, 10, and 11 from $K^{ji}$, respectively.

\item [{Step~3.2.4}] $V_{j}$ sends the encoded sequence of travel qubits $T^{ij}$ to $V_i$ after concatenating some Bell states and applying a permutation operator. The correct order of travel qubits to perform eavesdropping checking and obtaining $T^{ij}$ is revealed by $V_i$ only after he receives an authenticated acknowledgment from $V_i$. 

\item [{Step~3.2.5}] $V_i$ announces his random sequence $K^{ij}$ if they obtain error below the threshold value. $V_j$ reveals the permutation operation only after he gets to know $K^{ij}$. Subsequently, he performs a Bell measurement on the pair of home and travel qubits from $H^{ij}$ and $T^{ij}$ and obtains the random sequence $K^{ji}$ sent by $V_{j}$. The symmetric key between $V_{i}$ and $V_{j}$ is obtained as $K^{ij}\oplus K^{ji}$.

\end{description}

Another modification of QAV-1 allows semiquantum users, i.e., voters with limited quantum resources, to perform QAV.
Specifically, a semiquantum or classical voter is defined as the one who can (1) measure the quantum state in the computational basis only, (2) prepare the quantum state in the computational basis only, and (3) do nothing and/or reflect a quantum state which is sent to him by a quantum user. 
The steps involved in semiquantum AV protocol (QAV-4) with classical voters inspired from semi-QKD protocol \cite{krawec2015mediated} can be described as follows:

\begin{description}

\item [{Step~4.2.1}] CA generates $l$ Bell states $\vert \phi \rangle$ to be shared with voters $V_{i}$ and $V_{j\neq i}$. He forms two ordered sequences of all the first qubits $F^{ij}=\{f^{ij}_{l'} \}_{j\neq i}$ and second qubits $S^{ij}=\{s^{ij}_{l'} \}_{j\neq i}$ with $l'=0,1,\dots,l-1$. Finally, he sends sequences $F^{ij}$ and $S^{ij}$ to voters $V_{i}$ and $V_j$, respectively.

\item [{Step~4.2.2}] $V_{i}$ prepares a random string $R^{ij}=\{r^{ij}_{l'} \}_{j\neq i}$, where $r^{ij}_{l'}\in \{0,1 \}$.
He measures the qubits $f^{ij}_{l'}$ if $r^{ij}_{l'}=0$ and keeps it unchanged otherwise. 
He also notes the measurement outcomes in a string $T^{ij}=\{t^{ij}_{l'} \}_{j\neq i}\, \forall r^{ij}_{l'}=0$ and prepares fresh qubits $|t^{ij}_{l'}\rangle$, where $t^{ij}_{l'}\in \{0,1 \}$. He reinserts $|t^{ij}_{l'}\rangle\, \forall r^{ij}_{l'}=0$ in the remaining qubits of $F^{ij}$ and sends $F^{\prime ij}$ to CA.\\
Independently, $V_{j}$ adopts the same procedure to obtain $S^{\prime ij}$ and then sends it to CA.

\item [{Step~4.2.3}] CA performs the Bell measurement on the respective pairs in sequences $F^{\prime ij}$ and $S^{\prime ij}$ and records $C^{ij}=\{c^{ij}_{l'} \}_{j\neq i}$, where $c^{ij}_{l'}=0$ if the measurement outcome is $\vert \phi \rangle$ and $c^{ij}_{l'}=1$ otherwise. Finally, he announces $C^{ij}$.

\item [{Step~4.2.4}] $V_{i}$ and $V_j$ announce their random strings $R^{ij}$ and $R^{ji}$, respectively. They obtain the error rate in cases $r^{ij}_{l'}=1=r^{ji}_{l'}$ as CA would have announced $c^{ij}_{l'}=0$ ideally. If the error is below the threshold value, they obtain $\bar{K}^{ij}$ and $\bar{K}^{ji}$ (approximately of size $l/4$) as the subset of $T^{ij}$ and $T^{ji}$ for the cases when $r^{ij}_{l'}=1=r^{ji}_{l'},\, c^{ij}_{l'}=0$.

\item [{Step~4.2.5}] Ideally, $\bar{K}^{ij}=\bar{K}^{ji}=K^{ij}$, otherwise $V_{i}$ and $V_j$ may perform post-processing of the key to obtain symmetric key.

\end{description}

{QAV-3 (QAV-4) has the same arrangement of voters in AV net as QAV-1 (QAV-2).}

\subsection{Iterative QAV protocols}

In such class of protocols, we will be using Bell states and their quantum correlations. We assume that we have a quantum voting network with a {semi-honest} central authority CA and $n$ voters $V_i$.

\subsubsection{QAV-5: Bell state based iterative probabilistic QAV protocol}

In fact, QAV-2 proposed earlier can be implemented in an iterative manner to give us a probabilistic outcome of the AV. Thus, the arrangement of voters has the same graph structure as in QAV-2.
The steps involved in this protocol are as follows:

\begin{description}

\item [{Step~5.1}-{Step~5.7}] CA performs QAV-2 with all the voters for $l=1$. Thus, $\mathcal{V}_n=1 \Leftrightarrow R=1$ and leads to an inconclusive result otherwise.

\end{description}

All the parties repeat the protocol an arbitrary number of time $l'$ until they either get $R=1$ or conclude with probability $\left(1-2^{-l'}\right)$ that $\mathcal{V}_n=0$. Though the scheme remains probabilistic as is QAV-2, but it requires $^n C_{2}\times l'$ copies of Bell states which will be less than that in QAV-2 if  $R=1$ is obtained in $l'<l$, where $l$ is a constant number of Bell states used in QAV-2. 

\subsubsection{QAV-6: Bell state based iterative QAV protocol}

Let us now present another protocol which where the arrangement of voters is in circular order as shown in Fig. \ref{fig1} (b). Here, in each iteration the CA will  generate one Bell state, keep one particle with himself while the other particle travels through each of the voters and finally comes back to CA. The main advantage of this scheme is that the voting process requires less than $1+\log_2 n$ copies of Bell states. The steps involved in the protocol (QAV-6) are as follows: 

\begin{description}

\item  [{Step~6.1}] Same as Step 1.1 of QAV-1. 

\item [{Step~6.2}] CA generates a Bell state $\vert \phi \rangle $. CA sends the second qubit of $\vert \phi \rangle $ to $V_0$ as travel qubits in a secure manner. He keeps the first qubit as home qubits with himself.

\item [{Step~6.3}] After ensuring that there is no eavesdropping attempt, $V_0$ applies $\sigma_z$ operation to the travel qubit to veto the proposal and does nothing in case he supports the proposal.

\item [{Step~6.4}] $V_0$ sends the encoded travel qubit to $V_1$ in a secure manner, who encodes his message in the same way as  $V_0$.\\
Voter $V_i \,\forall\, 0\leq i \leq n$ receives the travel qubits from $V_{i-1}$  and sends it to $V_{i+1}$ after applying $\sigma_z$ operation to veto the proposal.

\item [{Step~6.5}] $V_n$ receives the travel qubits from $V_{n-1}$ in a secure manner and encodes his vote. Finally, he sends the travel qubits to CA in a secure manner.

\item [{Step~6.6}] CA measures the final state $\vert \phi' \rangle =\sigma_z^{k} \vert \phi \rangle $, if $k$ voters vetoed the proposal, in the Bell basis. If $\langle\phi' | \phi\rangle=0$, CA announces $\mathcal{V}_n=1$, while $\langle\phi' | \phi\rangle=1$ leads to an inconclusive outcome.

\item [{Step~6.7}] In case of an inconclusive outcome, CA repeats Steps 6.2-6.6 with the $k$ voters applying unitary $ \sigma_{z}(t) = \begin{pmatrix}
1 & 0 \\
0 & \exp\left({i \pi}{2^{-t}}\right)
\end{pmatrix}$ 
on the travel qubit to veto the proposal in the $t$th iteration.\\
In each iteration, if CA gets $\langle\phi' | \phi\rangle=0$ in the measurement outcome he announces $\mathcal{V}_n=1$,
while $\langle\phi' | \phi\rangle=1$ leads to an inconclusive outcome as $2^{-t}k$ is even (including zero).\\

\item [{Step~6.7}] CA repeats Step 6.7 until he gets $\mathcal{V}_n=1$ or gets $\mathcal{V}_n=0$ conclusively. It should take at most $1+\log_2 n$ number of iterations to yield a conclusive outcome for $n$ voters.

\end{description}

\subsection{QAV-7: Deterministic QAV protocol}

Finally, we propose a QAV protocol that can succeed with unit probability in a single iteration. This circular scheme (cf. Fig. \ref{fig1} (b)) is based on mulitiparty densecoding \cite{banerjee2018qc}. Here, every Voter $V_i$ is assigned a subgroup $g_{i}=\left\{I, O_{i} \right\} $ of a group of operations $G_{2^m}$ to encode his information, where $V_i$ applies $O_{i}$ to veto the proposal while does nothing otherwise.  The group $G_{2^m}=\{I,\sigma_x, i\sigma_y, \sigma_z\}^{\otimes \lceil \log_2 m\rceil}$ with at least ${2^m}$ elements is obtained from the modified Pauli group (an Abelian group under multiplication obtained by neglecting the global phase of the states post-operation). The ${2^m}$ elements of the group $G_{2^m}$ generate quantum states mutually orthogonal to each other enabling it useful for multiparty densecoding (see \cite{banerjee2018qc} for detail). The subgroups assigned for encoding are pairwise disjoint $g_{i}\cap g_{j}=\left\{ {I}\right\} \forall i,j\in\{0,1,\dots,n-1\}$ and $O_{0}O_{1}\cdots O_{n-1}=I$. A few examples of the operations $\{g_{i} \}_{i=0,1,\dots,n-1}$ for different values of $n$ are given in Table \ref{tab:ex}. 
In principle, the assignment of subgroups to voters for encoding can be performed randomly as distribution of secret index in \cite{wang2016qv}, which forbids some participants with the help of CA to identify the voters vetoing the proposal. Thus, each voter knows only the subgroup assigned to him.

\begin{table}
\caption{We present some examples of the quantum states and corresponding quantum operations required for QAV-7.} \label{tab:ex}
\centering
\begin{tabular}{ccc}
\noalign{\smallskip}\hline
\bfseries Number of voters  & \bfseries Quantum state    & \bfseries Operation $O_i$ of voters $V_i$ used for vetoing \\
\noalign{\smallskip}\hline
3    & Bell or GHZ state  & $ O_{0}=X ,\,O_{1}=iY,\,O_{2}=Z  $\\

4  & GHZ state & $ O_{0}=X\otimes{I} ,\,O_{1}=iX\otimes X,\, O_{2}=iY\otimes X ,\,O_{3}=iY\otimes{I}$\\

4  & 4-qubit cluster state  & $ O_{0}=X\otimes iY ,\,O_{1}=X\otimes Z ,\, O_{2}=iY\otimes Z ,\,O_{3}=iY\otimes iY $ \\
\noalign{\smallskip}\hline
\end{tabular}
\end{table}

The steps involved in this protocol (QAV-7) are as follows:

\begin{description}
\item [{Step~7.1}] Same as Step 1.1 of QAV-1. 

\item [{Step~7.2}] CA prepares an $m$-qubit entangled state $|\psi_{\rm in}\rangle$ (with $m\ge\left(n-1\right)$).

\item [{Step~7.3}] CA  prepares string of $l$ qubits ($l<m$) of $|\psi_0\rangle$ to send to $V_0$ as travel qubits in a secure manner. He keeps the string of the rest of the $m-l$ qubits as home qubits with himself.

\item [{Step~7.4}] After ensuring that there is no eavesdropping attempt, Voter $V_0$ encodes his vote.
Specifically, Voter $V_0$ applies operation $O_{0}$ on {all} the travel qubits to veto while does nothing to support the proposal. The operation of $V_0$ transforms the initial state $|\psi_{\rm in}\rangle$ to $|\psi_{0}\rangle$.

\item [{Step~7.5}] Voter $V_0$ sends the $l$ travel qubits of $|\psi_0\rangle$ to $V_1$ in a secure manner, who encodes his message using $g_{1}$ to transform the state to $|\psi_1\rangle$. 

\item [{Step~7.6}] Voter $V_i \,\forall\, 1\leq i \leq n-1$ receives the travel qubits of $|\psi_{i-1}\rangle$ from $V_{i-1}$  and sends the travel qubits of $|\psi_{i}\rangle$ to $V_{i+1}$ after encoding his message using $g_{i}$.

\item [{Step~7.7}] Voter $V_n$ receives the travel qubits of $|\psi_{n-1}\rangle$ from $V_{n-1}$ in a secure manner. He encodes his message using $g_{n}$ to obtain $|\psi_{n}\rangle$. Finally, he sends the travel qubits to CA in a secure manner.

\item [{Step~7.8}] CA measures $|\psi_{n}\rangle$ in the same basis he has prepared the initial state $|\psi_{\rm in}\rangle$. 
If $\langle\psi_{\rm in} |\psi_{n}\rangle=0$, CA announces $\mathcal{C}_n=1$ while he announces $\mathcal{C}_n=0$
for the measurement outcome $\langle\psi_{\rm in} |\psi_{n}\rangle=1$.

\end{description}

CA's announcement $\mathcal{C}_n=1$ corresponds to the situation $k\neq\{0,n\}$, i.e., at least one of the voters has vetoed the proposal. On the other hand, CA's announcement $\mathcal{C}_n=0$ corresponds to unanimity in the decision (all have either vetoed or not vetoed), i.e., $k=\{0,n\}$. 
Only the voters know their individual voting preferences, thus they can deduce whether the proposal is vetoed or not, i.e., $\mathcal{V}_n=0$ or 1, respectively.

\section{Security and efficiency analysis of the proposed scheme}\label{sec:SE}

A QAV scheme is expected to satisfy a few criteria of security listed in Section \ref{sec:basic}. Further, we may note that a QAV protocol is $\epsilon$-secure if it is $\epsilon$-indistinguishable from a perfectly secure (hypothetical) ideal QAV scheme following those listed conditions \cite{renner2008,muller2009composability}.
 In the following, we will explicitly show the security of our proposed schemes with regards to requirements for AV along these lines (which is summarized in Table \ref{tab:vp}).

\subsection{Eligibility}
In all the protocols, we are using the scheme of quantum digital signatures for authentication of the voters irrespective of whether CA performs the authentication or the voters authenticate each other among themselves. In this way, only the eligible voters will be allowed to vote and thus the eligibility condition is satisfied for all the proposed protocols.

\begin{table}
\caption{Comparison of the security of the proposed protocols with the existing schemes.  The asterisk in the column for the correctness corresponds to probabilistic nature of the scheme.} \label{tab:vp}
\centering
\begin{tabular}{ccccccc}
\noalign{\smallskip}\hline
\bfseries Protocol & \bfseries Eligibility & \bfseries Privacy & \bfseries Binding & \bfseries Verifiability & \bfseries Correctness & \bfseries Robustness \\
\noalign{\smallskip}\hline
RGQAV & $\times$ & $\checkmark$ & $\times$ & $\times$ & $\checkmark$ & $\times$\\
WQAV  & $\checkmark$ & $\checkmark$ & $\checkmark$ & $\checkmark$ & $\checkmark^*$ & $\times$ \\
QAV-1  & $\checkmark$ & $\checkmark$ & $\checkmark$ & $\checkmark$ & $\checkmark^*$ & $\checkmark$\\
QAV-2  & $\checkmark$ & $\checkmark$ & $\checkmark$ & $\checkmark$ & $\checkmark^*$ & $\checkmark$ \\
QAV-3  & $\checkmark$ & $\checkmark$ & $\checkmark$ & $\checkmark$ & $\checkmark^*$ & $\checkmark$ \\
QAV-4  & $\checkmark$ & $\checkmark$ & $\checkmark$ & $\checkmark$ & $\checkmark^*$ & $\checkmark$ \\
QAV-5  & $\checkmark$ & $\checkmark$ & $\times$  & $\checkmark$ & $\checkmark^*$ & $\checkmark$\\
QAV-6  & $\checkmark$ & $\checkmark$ & $\times$  & $\checkmark$ & $\checkmark$ & $\checkmark$ \\
QAV-7  & $\checkmark$ & $\checkmark$ & $\checkmark$ & $\checkmark$ & $\checkmark$ & $\checkmark$\\
\noalign{\smallskip}\hline
\end{tabular}
\end{table}

\subsection{Privacy}

An eavesdropper attempts to access the information a voter is transmitting to CA. Her endeavor would result in a message encoded (by voter $V_{j}$) joint state shared among CA and Eve (before a measurement performed by Eve and/or CA) which can be described as 
\begin{equation}
\rho^{V_{j}E}= p_0 \rho_0^{V_{j}}\otimes \rho_0^{E} +(1-p_0) \rho_1^{V_{j}}\otimes \rho_1^{E}, 
\end{equation}
where $p_0$ is the probability that $V_{j}$ supports the proposal. Eve will further discriminate $\rho_j^{E}$ to identify the secret value of $j$. However, the legitimate parties, i.e., voters and CA, would desire to adopt quantum cryptography tools, such as decoy state technique, to obtain the joint state in ideal situation as
\begin{equation}
\rho^{V_{j}E}_{\rm ideal}= \sum_i p_j \rho_i^{V_{j}}\otimes \rho^{E}, 
\end{equation}
which ensures that Eve has no information about the choice of the voter. Thus, $\epsilon$-privacy of a QAV scheme can be defined in the information theoretic description of security \cite{renner2008,muller2009composability} as $\min\limits_j \frac{1}{2}\vert\vert \rho^{V_{j}E}- \rho^{V_{j}E}_{\rm ideal}\vert\vert\leq \epsilon$. 
Here, we provide privacy of the voters for the proposed QAV schemes against some of the well-known individual attacks by an adversary as well as the collusion attacks by the legitimate parties. 

To begin with, we consider the intercept and resend attack by a non-participant Eve. In the intercept and resend attack, Eve intercepts the travel particles from one legitimate user to another. Subsequently, Eve prepares a random state (known to him) and sends it to the party who was intended to receive the intercepted particles. 
For example, in QAV-2, Eve may perform this attack by intercepting the $l$ copies of Bell state to be shared between all the pairs of voters by CA. She will be able to get the information about the shared symmetric keys used by the voter to cast their votes by sending the symmetric separable single qubit strings to both the voters. This will eventually give her access to all the secret information that voters were sharing. To prevent this attack, we can employ decoy qubit based eavesdropping checking (cf. Section \ref{sec:basic}), e.g., using the BB84 states ($\vert 0 \rangle, \vert 1\rangle, \vert + \rangle, \vert - \rangle$). Suppose $l$ Bell states are to be securely distributed between a pair of voters by CA who inserts $2l$ decoy states randomly before sending. The pair of voters measure the $2l$ decoy qubits to obtain the error rate and attribute all these errors to the eavesdropping attempts. Since Eve is ignorant about the position of the decoy states as well as the choice of randomly used basis for preparation of the decoy states so the voters will detect the presence of Eve by comparison of the measurement outcomes with that of the prepared state. This allows the voters to obtain the bounds on the information accessible to Eve on the remaining $l$-bits key they obtain eventually. The probability to detect the presence of Eve is given by $1-\frac{1}{4^{l/2}}$. Similarly, in QAV-1, the decoy state based eavesdropping checking technique is effective to circumvent the intercept and resend attack as it is an integral part of the QKA/QKD protocols used in the generation of symmetric keys between the voters. Along the same lines, all the proposed protocols are free from the intercept and resend attack by using the decoy qubit based eavesdropping checking while transmission of the qubits between two parties. 

Another type of attack strategy is entangle and measure attack. In such type of attacks, Eve entangles her ancilla  qubit with the travel qubit and measures her ancilla afterwards to get the information transmitted. {Specifically}, Eve prepares an ancilla  qubit in a superposition state as $\vert q \rangle_{E}= \alpha \vert 0 \rangle_E + \beta \vert 1 \rangle_E$ and then entangles it with the travel qubit using the CNOT gate with control on the ancilla and target on the travel qubit. It can be observed that the use of decoy qubits prepared in the BB84 states ($\vert 0 \rangle, \vert 1\rangle, \vert + \rangle, \vert - \rangle$) will result in the successful detection of Eve with success probability $\vert\beta\vert^2$ if Eve attacks the decoy states $\vert 0 \rangle$ and $\vert 1 \rangle$ while the state remains separable for the rest of the decoy states ($\vert + \rangle, \vert - \rangle$). Thus, the average probability of detecting Eve can be obtained as $\frac{\vert\beta\vert^2}{2}$ assuming that all decoy states are prepared with an equal probability. Notice that if Eve prepares ancilla with $\beta\rightarrow 0$ then the detection probability of Eve will be vanishingly small as in that case Eve neither disturbs the decoy qubit nor gains any information. 

Another significant attack is the man in the middle attack where Eve impersonates {as} a legitimate party. This attack can be prevented by the use of a secure authentication scheme \cite{qauthentication1,qauthentication2,qauthentication3} before sending of the actual sequence of particles. Further, we are using the quantum digital signatures which would protect us from this attack. 

Further, in a participant attack, a user or a group of users will either try to get some information about the voting pattern of the voters or try to influence the result of the voting without being detected. In QAV-1, QAV-3 and QAV-4, every voter $V_i$ generates a $l$-bit symmetric key with the rest of the voters using a QKA/QKD protocol, which is followed by an application of some logical operations of those keys before broadcasting the result. Thus, it is not possible for any voter to get the information about the voting pattern/preference of the other voters from the announced information. Similarly, it is applicable to the rest of the protocols, i.e., QAV-2, QAV-5, QAV-6 and QAV-7. 
However, in the collusion attack, $k<n$ voters out of the total $n$ voters collude to acquire the inaccessible information about the voting preferences of the rest of the $n-k$ voters and then try to change the outcome. In all of proposed protocols, we can see that it is not possible to violate the secrecy of the vote as well as the outcome of the voting process. For instance, QAV-7 is prone to the collusion attack by an arbitrary voter and CA as they know the choices by all voters in the end if the operations applied by the voters are public knowledge. Specifically, CA has the information of the final result after measurement and (all) the voter(s) have encoding operations, and thus together they have all the pieces required to get all the voting preferences, i.e., to identify the parties vetoing the proposal.
Here, this possibility is circumvented as the disjoint subgroups are assigned to every voter for voting in a random manner with neither CA nor the voters aware of the encoding operations used by the rest of the voters. 

We have shown here that the privacy of the votes can be accomplished against some of the popular outsider's and insider's attacks, but a more rigorous security proof against collective and coherent attacks will be performed in our future works.

\subsection{Binding}

 In all the protocols proposed here, an outsider (or a participant other than the voter) cannot change the vote encoded by any voter, and the same is already established in the context of privacy against denial of service and disturbance attacks. Further, in the probabilistic and deterministic protocols (i.e., QAV-1--QAV-4 and QAV-7), even the voter cannot alter the vote as they only get one chance to encode it, but in the iterative protocols, a dishonest voter may change his vote in every iteration, e.g, in QAV-4--QAV-5. However, the voter's change of the vote in the successive iterations neither allows him access to the partial tally of the votes nor compromises the privacy of the other voters. Thus, a voter cannot take advantage of changing the vote in every iteration to get a favourable final outcome of his choice. 

\subsection{Correctness}
The correctness of an $\epsilon$-correct QAV scheme requires that the result bit is generated wrong with probability $\rm{Pr}[\mathcal{W}_i=0\, \forall i\longrightarrow \mathcal{V}_n=1]\leq \epsilon $. The success  probability of  probabilistic protocols is given by $\frac{1}{2^l}$ where $l$ represents the number of bits used by each voter. So, probabilistic QAV protocols are $\epsilon$-correct with $\epsilon \geq  1-\frac{1}{2^l}$. In comparison to the probabilistic protocols, the iterative and deterministic QAV protocols can be implemented with a relatively small value $\epsilon$.  

\subsection{Verifiability}

The AV scheme is $\epsilon$-verifiable if every voter can confirm his vote with a probability of failing verifiability $\rm{Pr}[\mathcal{W}_i=j \longrightarrow \mathcal{V}_n=j\oplus 1]\leq \epsilon$. Notice that a voter (say $V_i$) can verify his veto ideally with unit probability, while any other voter may independently have supported the proposal which reduces the verifiability of the scheme as the voters supporting the proposal with input $\mathcal{W}_i=0$ would not be able to verify the outcome. Thus, as long as the scheme is $\epsilon_c$-correct it will lead to $\epsilon$-verifiability $\left(\epsilon>\epsilon_c\right)$. In our case, a party who vetoed the proposal can verify the outcome with unit probability in case of iterative (QAV-6) and deterministic (QAV-7) schemes. However, in case of probabilistic schemes, he will be able to verify the result as long as the correctness is ensured. Further, in case when the parties support the proposal, it does not appear possible to ensure verifiability without disclosing individual choices.

\subsection{Robustness}
Decoherence is the major challenge in the implementation of quantum communication. In the absence of an adversary, an interaction of the qubits with the ambient environment is expected to reduce the correctness by leading to a wrong outcome.
Any realistic physical implementation of the proposed protocols will always be noisy due to the presence of the surrounding environment. Further, the protocol will be practically useful only if it gives the correct result even in the presence of a limited amount of noise. Here, we will be comparing the feasibility of  the proposed protocols under the presence of noise by considering that the noise affects the qubits only when they travel from one party to the other. Further, we assume that the qubits that do not travel are hardly affected by the noise. In quantum information theory, the effect of noise on the quantum state $\rho_i$ evolving to $\rho_f$ is described as an operator-sum representation in terms of Kraus operators as \cite{kraus,nielsen} 
\begin{eqnarray}\label{Kr}
\rho_f= \sum _{i} E_{i}\rho_i E_{i}^{\dagger},\label{eq:Kr}
\end{eqnarray}
where $E_{i}$s are the Kraus operators with $\sum_{i}E_{i}^{\dagger} E_{i} = I$. 

To discuss the robustness of the proposed schemes, we study the effect of two of the most important noise channels, namely amplitude damping and phase damping, on the proposed protocols. The Kraus operators for amplitude damping are
\begin{equation}\label{AD-Kr}
E^{\rm AD}_0 = \left(
     \begin{array}{cc}
      1 & 0  \\
       0 & \sqrt{1-\eta_{a}}    
     \end{array}
   \right) \quad {\rm and} \quad
   E^{\rm AD}_1 = \left(
     \begin{array}{cc}
      0 & \sqrt{\eta_{a}} \\
       0 & 0     
     \end{array}
   \right)
\end{equation}
and those for phase damping are 
\begin{equation}\label{PD-Kr}
E^{\rm PD}_0 = \left(
     \begin{array}{cc}
      1 & 0  \\
       0 & \sqrt{1-\eta_{p}}    
     \end{array}
   \right)\quad {\rm and} \quad
   E^{\rm PD}_1 = \left(
     \begin{array}{cc}
      1 & 0 \\
       0 & \sqrt{\eta_{p}}     
     \end{array}
   \right).
\end{equation} 
These operators can be  substituted in Eq. (\ref{Kr}) to give us the final state with $\eta_{j}$ as the damping parameter.

Suppose an $n$ qubit initial pure state $\rho_i= \vert \Phi \rangle\langle \Phi \vert$ is used for the implementation of a protocol, with $m\, \left( n-m\right)$ home (travel) qubits denoted by $h  \, \left( t\right)$, then the final state before measurement can be written as
\begin{equation}
\rho_f^k= \sum _{i_j} \left\{I^{\otimes m}_h \otimes \left( E^{k}_{i_1}\otimes \dots  E^{k}_{i_j} \dots \otimes E^{k}_{i_{n-m}}\right)_t\right\} \: \rho_i \:\left\{I^{\otimes n}_h \otimes \left( E^{k}_{i_1}\otimes \dots  E^{k}_{i_j} \dots \otimes E^{k}_{i_{n-m}}\right)_t \right\}^{\dagger},
\end{equation}
where $ E^{k}_{i_j}$ are the Kraus opertors of amplitude or phase damping with $k\in \{\rm{AD,PD}\}$. The effect of the noise can be quantified by a distance based measure, known as the square of fidelity (henceforth referred to as fidelity), given by
\begin{equation}
F^k= \langle \Phi^{f} \vert	\rho_f^k \vert \Phi^{f} \rangle,
\end{equation}
where $\vert \Phi^{f} \rangle$ represents the final state that the initial pure state $\vert \Phi \rangle$ should have been after performing all the encoding operations by every party in a decoherence free environment. In our case, we consider that all the encoding operations are equi-probable and hence calculate the average fidelity for each of the proposed QAV protocols. Before we proceed further, notice that the average fidelity quantifies the robustness of the scheme as the low fidelity corresponds to the wrong outcome.

\begin{figure}[tb]
\includegraphics[width=0.3\textwidth]{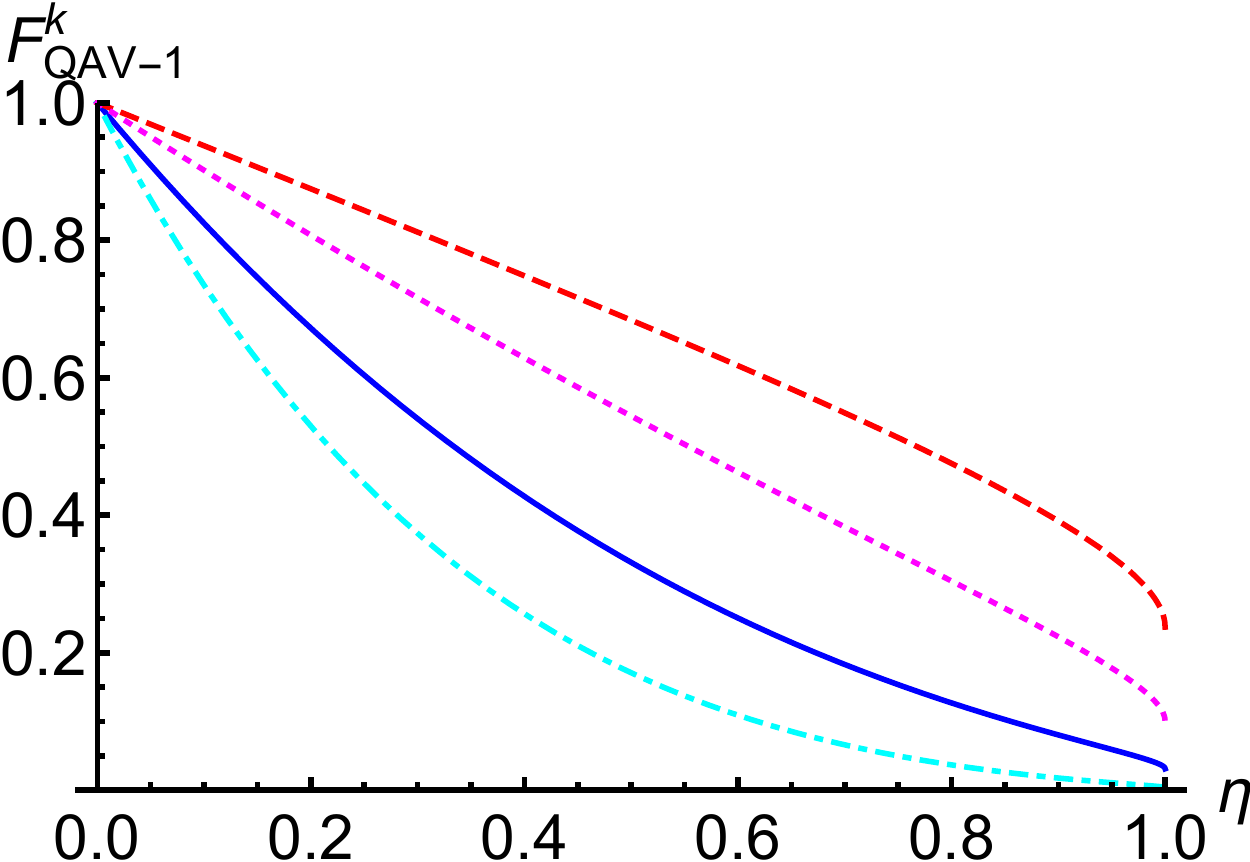}\, \includegraphics[width=0.3\textwidth]{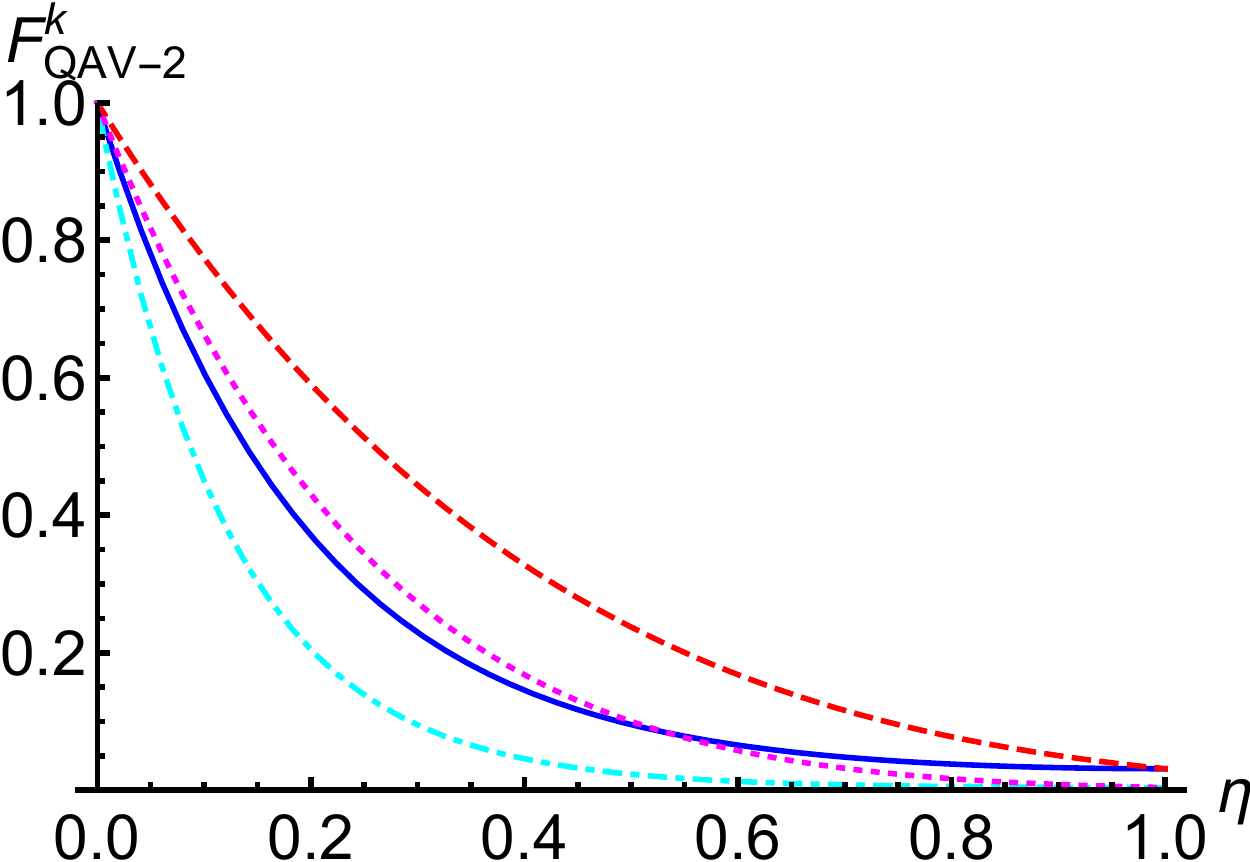}\, \includegraphics[width=0.3\textwidth]{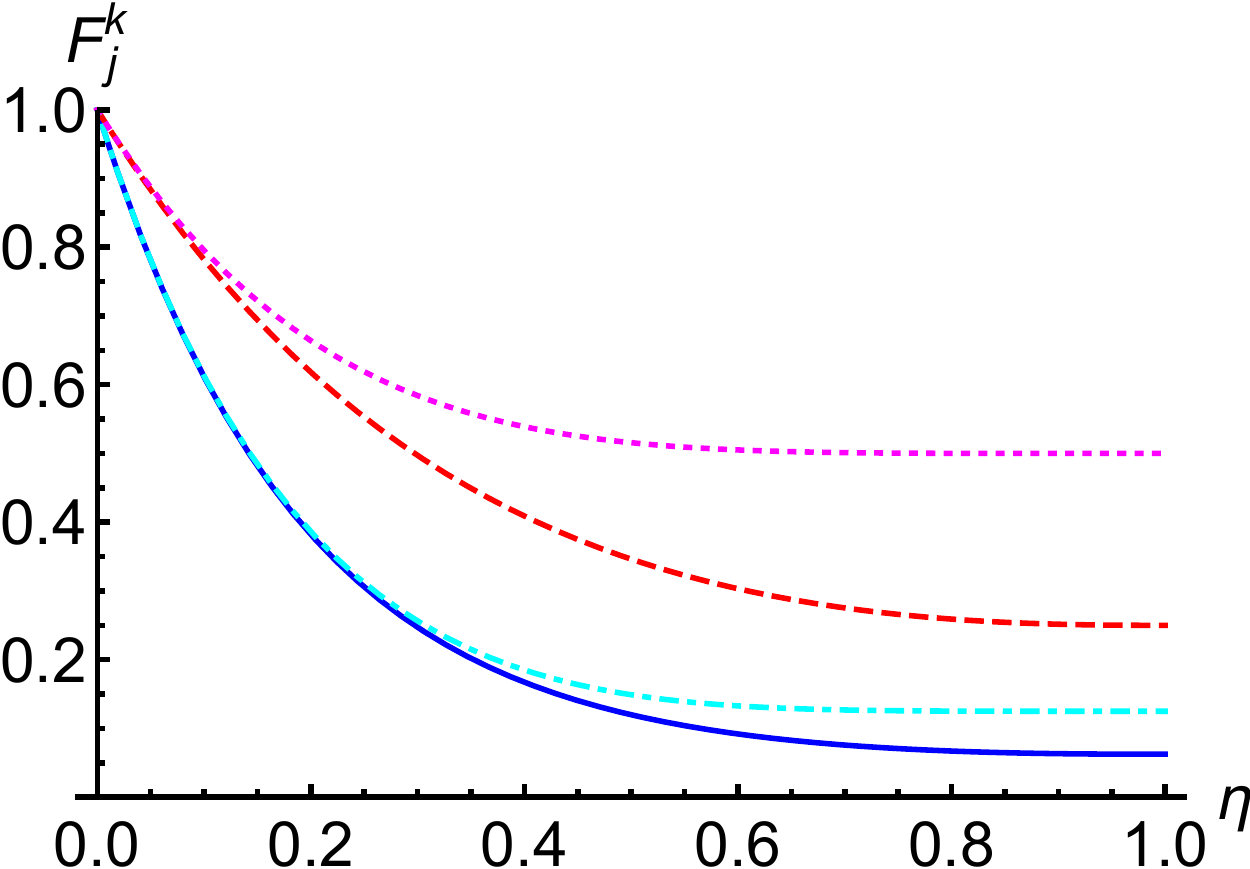}
 \centerline{ \small (a) \hspace{.3\hsize} (b)  \hspace{.3\hsize} (c)}
\caption{(Color online) Variation of average fidelity for (a) QAV-1, (b) QAV-2 and (c) QAV-6 and QAV-7  with damping factor of the amplitude damping (in the smooth (blue) and dot-dashed (cyan) lines) and phase damping (in the dashed (red) and dotted (magenta) lines) channels with $k\in \{\rm{AD,PD}\}$. In (a)-(b), the smooth (blue) and red (dashed) lines (the cyan (dot-dashed) and magenta (dotted) lines) correspond to QAV with 5 (8) {key bits size}. In (c), the smooth (blue) and red (dashed) lines (the cyan (dot-dashed) and magenta (dotted) lines) correspond to $j$ as QAV-6 (QAV-7) with 4 voters.}  
\label{fig2}
\end{figure}

The QAV-1 protocol is dependent upon the choice of QKA/QKD protocol used for the initial key generation. Without loss of generality, we consider BB84 protocol in this case for the analysis. It involves the sending of BB84 states from one voter to the other voters for creating a $l$-bit keys between every pair of voters. The average fidelity for generation of $l$-bit keys for every pair of voters under amplitude damping noise is computed to be $\frac{1}{4^l} \left(\sqrt{1-\eta_a }-\eta_a +3 \right)^l $ while under phase damping noise it is found to be $\frac{1}{4^l} \left(\sqrt{1-\eta_p }+3 \right)^l $. Thus, the fidelity depends on the noise parameter values as well as the number of key bits required for working of the protocol as can be seen from Fig. \ref{fig2} (a). Specifically, the protocol is robust for the small values of noise parameters ($\eta_p$ or $\eta_a$), and a higher value of noise reduces the fidelity significantly and thus rendering the protocol practically ineffective.  Since QAV-1 is a probabilistic AV protocol and for $l=10$ we get a conclusive outcome with probability 99.9\%, which can be further improved by increasing the number of key bits $l$. However, with an increase in the key size the robustness decreases and thus a trade-off between correctness and robustness of the probabilistic QAV schemes is observed. Further, we can observe that the amplitude damping noise has a greater impact on the average fidelity in comparison to that for the phase damping noise due to the presence of fast decaying term $-\eta_a$ in the former case. Similarly, in QAV-2 protocol based on the Bell states shared among two voters, the average fidelity for generation of $l$-bit keys among every pair of voters under amplitude damping noise is $\left(1+ \frac{1}{2} (\eta_a -2) \eta_a \right)^l$ while under phase damping noise is $(1 - \frac{\eta_p}{2})^l $. Interestingly, as QAV-5 protocol is similar to QAV-2 (with differences in the encoding and measurement stages), the average fidelity is the same as that for QAV-2. Along the same lines, the average fidelity for QAV-3 (orthogonal state based protocol) under amplitude damping noise is found to be $(1 - \frac{\eta_a}{2})^{l}$ while under phase damping noise it is computed as $(1 - \frac{\eta_p}{2})^{l/2} $ for only even values of $l$.  Further, average fidelity  for QAV-4 (semi-quantum protocol)  is obtained to be the same as QAV-2 as the communication complexity is same in both the schemes. Among these schemes, QAV-2 (and QAV-4 and QAV-5, too) is the least robust against noise (cf. Fig. \ref{fig2} (b)).

We further obtain the average fidelity of the transmitted states in QAV-6 and QAV-7 implemented by the four voters with the help of CA as  
\begin{eqnarray}
F^{\rm AD}_{\rm QAV-6} &=& -\frac{\eta_a ^5}{4}+\frac{5 \eta_a ^4}{4}-\frac{5 \eta_a ^3}{2}+\frac{1}{2} \sqrt{1-\eta_a } \eta_a ^2+\frac{5 \eta_a ^2}{2}-\sqrt{1-\eta_a } \eta_a -\frac{5 \eta_a }{4}+\frac{\sqrt{1-\eta_a }}{2}+\frac{1}{2} , \\ \nonumber
F^{\rm PD}_{\rm QAV-6} &=&  \frac{1}{2} \sqrt{1-\eta_p } \eta_p ^2-\sqrt{1-\eta_p } \eta_p +\frac{\sqrt{1-\eta_p }}{2}+\frac{1}{2} , \\ \nonumber
F^{\rm AD}_{\rm QAV-7} &=&  \frac{\eta_a ^{10}}{4}-\frac{19 \eta_a ^9}{8}+10 \eta_a ^8-\frac{197 \eta_a ^7}{8}+\frac{315 \eta_a ^6}{8}-\frac{349 \eta_a ^5}{8}+\frac{289 \eta_a ^4}{8}-\frac{195 \eta_a ^3}{8}+\frac{107 \eta_a ^2}{8}-5 \eta_a +1, \\ \nonumber
F^{\rm PD}_{\rm QAV-7} &=& -\frac{\eta_p ^5}{2}+\frac{5 \eta_p ^4}{2}-5 \eta_p ^3+5 \eta_p ^2-\frac{5 \eta_p }{2}+1,
\end{eqnarray}
respectively.
In QAV-6, one of the qubits of the Bell states is transmitted five times through the noisy environment. Therefore, the effect of amplitude damping is more severe than that of phase damping. In QAV-7 protocol, a deterministic scheme among the four voters with two travel qubits has twice more travel qubits than that in QAV-6. The expressions for average fidelity are along the expected lines with amplitude damping having more adverse effect. Fig. \ref{fig2} (c) shows a comparison of average fidelity for QAV-6 and QAV-7 for the case of four voters. We can see that the robustness of the protocol is dependent upon the noise parameters. In the case of practical implementation, all the protocols may be observed robust up to moderate decoherence rates and the robustness decrease as the noise parameters increase.

\subsection{Efficiency of the protocols}

The performance of a quantum communication scheme can be quantified in terms of qubit efficiency, given by \cite{cabello2000} 
\begin{equation}
\eta = \frac{c}{q+b},
\end{equation}
where $c$ is the number of classical bits transmitted, $q$ is the minimum number of qubits required, while $b$ is the additional classical bits of information required for secure transmission. It is to be noted here that we do not consider the classical bits exchanged during eavesdropping checking while computing $\eta$. {Further, the number of qubits required can be written as $q=Q+\delta t$, where $Q$ represents the  total $Q$ qubits used in the protocol, while $t$ represents the number of travel qubits in the corresponding protocol. The factor of $\delta \neq 0$ is decided to achieve the desired level of security of $t$ travel qubits by using $\delta t$ decoy qubits. In QAV protocols, $c=1$ as we require only one bit of information $\mathcal{V}_n$ after the completion of the protocol. Let us now compare the efficiency of the existing QAV protocols along with that of our proposed QAV protocols.  

To begin with, let us look at the efficiency of WQAV protocol. In this protocol, CA has to establish a $l$ qubit key with all the $n$ voters using BB84 protocol, which requires the exchange of a minimum $4nl$ qubits. Thereafter, CA would share $(1+\delta_0)l$ ordered copies of $n$-qubit GHZ state with the voters, which will require an additional $nl(1+\delta_0)\delta_1$ decoy qubits. Here, $\delta_0$ and $\delta_1$ are the security parameters  for checking the GHZ correlations and eavesdropping checking, respectively. Thus, $q=nl(5+\delta_0+\delta_1+\delta_0\delta_1)$. 
The voters further require an exchange of a total of $b=nl$ classical bits to CA, and hence the efficiency is given by $\{nl(6+\delta_0+\delta_1+\delta_0\delta_1)\}^{-1}$. Though a detailed security of the RKQAV protocol was not reported we can calculate its qubit efficiency in the similar manner to that of WQAV protocol. This also requires the transfer of $(1+\delta_0)nl$ GHZ particles (qubits) from CA to the $n$ voters. After preforming some operations on their GHZ particles, the voters will then return back a total $nl$ particles to CA. To ensure the detection of Eve during transfer of qubits, we require additional $nl(2+\delta_0)\delta_1$ decoy qubits. Thus, qubit efficiency is  $\{nl(1+\delta_0+2\delta_1+\delta_0\delta_1)\}^{-1}$ as $b=0$.

\begin{table}
\caption{Comparison of qubit efficiency for the existing as well as the proposed protocols.} \label{tab:qe}
\centering
\begin{tabular}{cccc}
\noalign{\smallskip}\hline
\bfseries Protocol & \bfseries Quantum state used & \bfseries Qubit efficiency ($\eta$) & \bfseries $\eta$  for 4 voters \\
\noalign{\smallskip}\hline
RGQAV & n-party GHZ states & $\{nl(1+\delta_0+2\delta_1+\delta_0\delta_1)\}^{-1}$ & $\frac{1}{200}$\\
WQAV & n-party GHZ states & $\{nl(6+\delta_0+\delta_1+\delta_0\delta_1)\}^{-1}$  & $\frac{1}{360}$\\
QAV1 & Based on QKA/QKD scheme used & $\{(2n-1))nl \}^{-1}$(BB84 based) & $\frac{1}{280}$\\
QAV2 & Bell states & $\{((n-1)(\delta_1+1)+1) nl\}^{-1}$ & $\frac{1}{280}$\\
QAV3 & Bell states & $\{(\frac{(n-1)(\delta_1+1)}{2}+4)nl\}^{-1}$ &  $\frac{1}{280}$\\
QAV4 & Bell states & $\{nl(4n-3)\}^{-1}$ & $\frac{1}{520}$ \\
QAV5 & Bell states & $\{((n-1)(\delta_1+1)+1) nl\}^{-1}$ & $\frac{1}{280}$\\
QAV6 & Bell states & $\{((n+1)(1+\delta_1) +2)l \}^{-1}$ & $\frac{1}{24}$ \\
QAV7 & m-qubit entangled state with $m\geq (n-1)$ & $\{m + (n+1)(1+\delta_1)l +1\}^{-1}$ & $\frac{1}{24}$\\
\noalign{\smallskip}\hline
\end{tabular}
\end{table}

Similarly, we can compute the qubit efficiency of the proposed probabilistic QAV protocols. QAV-1 protocol is based on the generation of $l$ bit key among all pairs of $n$ voters using any of the QKD or QKA protocol. For instance, considering $l$ bit key shared among arbitrary two voters using the BB84 QKD protocol, which involves $q={^n C_2}\, 4l$. Further, after generation of the symmetric keys every voter has to publicly announce the $l$ bits of classical information, which makes $b=nl$ and the efficiency is calculated as  $\{(2n-1))nl \}^{-1}$. Similarly, QAV-2 requires the sharing of the $l$ Bell states, among all pairs of voters. The total number of qubits used are $q={^nC_2}\, 2l(\delta_1+1)$, and $nl$ classical bits are required.   Hence, the qubit efficiency of QAV-2 can be calculated as  $\{((n-1)(\delta_1+1)+1) nl\}^{-1}$. QAV-3 uses an orthogonal state based QKA to generate $l$ bit key between any pair of voters. The total number of qubits required $q={^nC_2}\, l(\delta_1+1)$ with $b=4nl$ classical bits are required. This results in qubit efficiency as $\{(\frac{(n-1)(\delta_1+1)}{2}+4)nl\}^{-1}$. In QAV-4, semi-QKD is employed by the parties which requires $q={^nC_2}\, 8l$ with $b=nl$ to generate $l$ bit keys. This leads to the efficiency of protocol as  $\{nl(4n-3)\}^{-1}$ by including classical communication post QKD step.
 
Along the same lines, the qubit efficiency of the proposed iterative QAV protocols can also be obtained.
The efficiency of protocol QAV-5 is similar to that of QAV-2. Let's now look at efficiency analysis of QAV-6. In this protocol a Bell state is generated and then one qubit is kept by the CA while the other qubit will be travelling among the $n$ voters for casting the vote and will return back to CA.  In this case, $q=((n+1)(1+\delta_1) +2)l$ and $c=0$ which leads to efficiency calculated as $\{((n+1)(1+\delta_1) +2)l \}^{-1}$. Here, $l$ refers to the number of iterations required to get a conclusive outcome and its maximum value is given by $1+\log_2n$. In QAV-7, we are using the dense coding scheme to arrive at the voting outcome. Here, CA generates a $m$-qubit entangled state and then $l$ qubits of that state are transferred to all the voters one by one  and finally returned back to CA which leads to $q=m+(n+1)l$. Finally the revealing of outcome results in use of $b=1$ classical bit of information which leads to efficiency as $\{m + (n+1)(1+\delta_1)l +1\}^{-1}$. The comparison of the efficiencies is presented in Table \ref{tab:qe}. Without loss of generality, we further calculated the efficiency in a special case of 4 voters. We can see that some of our proposed protocols fare better than the RKQAV and WQAV protocols. Interestingly, we can clearly observe that for 4-party voting example of all the mentioned protocols, QAV-6 and QAV-7 have the best efficiency. In fact, this is true for voting with higher number of voters too.

\section{Conclusions}\label{sec:conc}

Veto is a form of voting in which the proposals are accepted only in the case of consensus among the involved parties. Further, there is a heightened interest in designing protocols for secure anonymous veto using the quantum resources. In this study, we have proposed a number of quantum anonymous veto protocols based on various degrees of available quantum resources. In the present work, we have  classified the protocols based on the probabilistic, iterative and deterministic approaches in order to accomplish the task and arrive at the desired outcome. We have further explored the intrinsic connections between DC-net and AV-nets. We have performed a security and efficiency analysis  of the proposed protocols and established the proposed schemes are secure against some of the widely studied attacks. We have also performed a comparative analysis of the performance of the existing QAV schemes. In addition, we have examined the robustness of the proposed protocols under realistic physical systems, i.e., the effect of noise on the implementation. The analysis shows that the proposed schemes are robust in the presence of weak noise. Our comparison of the performance of the proposed schemes establishes that the deterministic QAV scheme (QAV-7) is an optimal protocol to accomplish the desired task. A bipartite entanglement QAV scheme (QAV-6) is also observed to be more efficient and robust than all the existing iterative and probabilistic QAV schemes. However, QAV6 does not satisfy the requirement of binding. Further, all the protocol proposed here can be experimentally implemented using the currently available technology. However, a particular laboratory or a company may have its own restrictions. For example, a laboratory may have capacity to produce the single qubit states only. Thus, in short, the set of protocols proposed here and the comparison tables reported here leads to an opportunity to different organizations having varied expertise and capability to implement QAV based on the available  resources and the exact requirement(s).   The recent application of AV schemes to perform sealed bid auction  \cite{bag2019seal} by performing AV for each bit of the placed bids starting from the most significant bit to the least significant bit will hopefully encourage the utilization of the proposed schemes for other socioeconomic tasks of relevance. We hope that the set of  proposed QAV schemes will motivate experimentalists to realize the protocols and find them useful in performing veto and auction in the real life situations.

\section*{Acknowledgement}
SM, APar and APat  acknowledge the support received through the  project “Partnership 2020: Leveraging US-India Cooperation in Higher Education to Harness Economic Opportunities and Innovation” which is enabling a collaboration between University of Nebraska at Omaha, and JIIT, Noida. They also acknowledge Deepak Khazanchi for his interest in this work. KT acknowledges GA \v{C}R (project No. 18-22102S) and support from ERDF/ESF project `Nanotechnologies for Future'
(CZ.02.1.01/0.0/0.0/16\_019/0000754). 



\end{document}